 \documentclass[final,5p,times,twocolumn,12pt,sort&compress]{elsarticle}
\usepackage{soul}
\usepackage{lineno,hyperref}
\usepackage{txfonts}
\usepackage{graphicx}
\usepackage{dcolumn}% Align table columns on decimal point
\usepackage{array}
\usepackage{color}
\usepackage{multicol}
\usepackage{amssymb}
\usepackage{amsmath}
\usepackage{type1cm}
\usepackage{caption}
\usepackage{natbib}
\usepackage{lineno,hyperref}
\usepackage{comment}
\usepackage{longtable}
\usepackage{ctable}
\modulolinenumbers[5]

\onecolumn

\journal{JQSRT, prepared by using elsarticle.cls}
\begin{document}
\begin{frontmatter}
\title{Extended calculations of energy levels, radiative properties,  and lifetimes for
P-like \mbox{Ge XVIII}}

\author[hb,fd]{Kai Wang}
\author[hb]{Xiao Han Zhang}
\author[fd]{Chun Yu Zhang}
\author[hb]{Wei Dang}
\author[hb]{Xiao Hui Zhao}
\author[nudt]{Zhan Bin Chen\corref{bin}}
\author[fd]{Ran Si\corref{ran}}
\author[fd]{Chong Yang Chen}
\author[bj]{Jun Yan}

\cortext[bin]{chenzb008@qq.com}
\cortext[ran]{rsi13@fudan.edu.cn}

\address[hb]{Hebei Key Lab of Optic-electronic Information and Materials, The College of Physics Science and Technology, Hebei University, Baoding 071002, China}
\address[fd]{Shanghai EBIT Lab, Key Laboratory of Nuclear Physics and Ion-beam Application, Institute of Modern Physics, Department of Nuclear Science and Technology, Fudan University, Shanghai 200433, China}
\address[nudt]{School of Science, Hunan University of Technology, Zhuzhou, 412007, China}
\address[bj]{Institute of Applied Physics and Computational Mathematics, Beijing 100088, China}

\begin{abstract}
Using the multiconfiguration Dirac-Hartree-Fock (MCDHF) and the relativistic configuration interaction (RCI) methods,
a consistent set of transition energies and radiative transition data for the lowest 150 states of the $3s^2 3p^3$, $3s 3p^4$, $3s^2 3p^2 3d$, $3s 3p^3 3d$, $3p^5$, and $3s^2 3p 3d^2$ configurations in  P-like \mbox{Ge XVIII} is provided.  
To assess the accuracy of the MCDHF transition energies, we have also performed calculations using the many-body perturbation theory (MBPT). 
Comparisons are made between the present MCDHF and MBPT data sets, as well as with other available experimental and theoretical values. 
The resulting accurate and consistent MCDHF data set will be useful for  line identification and modeling purposes. These data can be considered as a benchmark for other calculations.
\end{abstract}


\end{frontmatter}

\twocolumn
\section{Introduction}

Because  of its applications in fusion plasmas,  the phosphorus isoelectronic sequence from \mbox{Zn XVI} to \mbox{Kr XXII} has  received great attention~\cite{Shirai.2007.V36.p509,Saloman.2007.V36.p215,Chen.2002.V66.p133}. 
The \mbox{Zn XVI}, \mbox{Se XX}, and \mbox{Kr XXII} spectra were measured in tokamak plasmas~\cite{Wouters.1988.V5.p1520,Roberts.1987.V35.p2591}. 
Using the National Institute of Standards and Technology (NIST) electron beam ion trap (EBIT), the extreme-ultraviolet (EUV) spectra of  \mbox{Kr XXII} were observed~\cite{Podpaly.2014.V47.p95702}. 
As regards P-like \mbox{Ge XVIII}, the electric-dipole (E1) transition array $(1s^2 2s^2 2p^6) 3s^2 3p^3$ -- $3s^2 3p^2 3d$ was observed by ~\citet{Sugar.1991.V8.p22}. 
The magnetic dipole (M1) transitions within the $3s^2 3p^3$ configuration of  \mbox{Ge XVIII} were measured by \citet{Denne.1984.V1.p296} in a tokamak discharge. 
 
On the theoretical side,  excitation energies and  transition rates for the low-lying 41 states of the $3s^2 3p^3$, $3s 3p^4$, and $3s^2 3p^2 3d$  configurations in \mbox{Ge XVIII} were provided by different calculations~\cite{Hu.2020.V137.p1141,Vilkas.2004.V37.p4763,Charro.2000.V33.p1753,Fritzsche.1998.V68.p149,Huang.1984.V30.p313,Fischer.1986.V19.p137}. Atomic parameters for higher-lying levels of P-like ions, such as the $3s 3p^3 3d$   levels,  
are also needed for applications in plasma physics~\cite{Fawcett.1975.V65.p623,Fawcett.1972.V5.p1255,Fawcett.1972.V5.p2143,Wang.2018.V235.p27,Song.2020.V247.p70}. 

%Given this background, we carried out high-precision benchmark calculations of P-like \mbox{Zn XVI}~\cite{Wang.2018.V235.p27} by using a state-of-the-art method, namely, the multi-configuration Dirac-Hartree-Fock (MCDHF) method combined with the RCI approach~\citep{FroeseFischer.2016.V49.p182004}. 
The present study is a continuation of our recent work~\cite{Wang.2018.V235.p27,Song.2020.V247.p70} on P-like ions, in which a complete accurate data set of excitation energies and  radiative rates involving high-lying levels  in P-like \mbox{Ge XVIII} is provided. By using a state-of-the-art method, namely, the multi-configuration Dirac-Hartree-Fock (MCDHF) method combined with the relativistic configuration interaction (RCI) approach~\citep{FroeseFischer.2016.V49.p182004}, excitation energies, wavelengths, lifetimes, and radiative transition data including line strengths, oscillator strengths, and transition rates, are provided for the lowest 150 levels of the $3s^2 3p^3$, $3s 3p^4$, $3s^2 3p^2 3d$, $3s 3p^3 3d$, $3p^5$, and $3s^2 3p 3d^2$
%$3s^2 3p^3$, $3s 3p^4$, $3s^2 3p^2 3d$, $3p^5$, $3s 3p^3 3d$, $3s^2 3p 3d^2$, $3s^2 3d^3$,  $3p^4 3d$, $3s 3p^2 3d^2$,  $3p^3 3d^2$,  $3s 3p 3d^3$,  and $3s^2 3p^2 4s$ 
configurations. To assess the accuracy of the MCDHF transition energies, we have also performed  calculations and provided excitation energies for \mbox{Ge XVIII} using the many-body perturbation theory (MBPT)~\citep{Lindgren.1974.V7.p2441}.  This work extends and complements our long-term theoretical efforts~\citep{Wang.2014.V215.p26,Wang.2015.V218.p16,Wang.2016.V223.p3,Wang.2016.V226.p14,Wang.2017.V119.p189301,Wang.2017.V194.p108,Wang.2017.V187.p375,Wang.2017.V229.p37,Wang.2018.V235.p27,Wang.2018.V239.p30,Wang.2018.V234.p40,Wang.2018.V208.p134,Wang.2019.V236.p106586,Wang.2020.V246.p1,Chen.2017.V113.p258,Chen.2018.V206.p213,Guo.2015.V48.p144020,Guo.2016.V93.p12513,Si.2016.V227.p16,Si.2017.V189.p249,Si.2018.V239.p3,Zhao.2018.V119.p314} to provide atomic data for L- and M-shells systems with high accuracy.

The paper is organized as follows.
The MCDHF and MBPT calculations are outlined in Sec.~\ref{cal}. In Sec.~\ref{results} we present our numerical results and compare them with measured values and previous calculations. Sec.~\ref{conclusions} is a brief summary.

\section{Calculations}~\label{cal}
\subsection{MCDHF}\label{Sec:MCDHF}
The MCDHF method implemented in the GRASP2K code~\citep{Jonsson.2013.V184.p2197,Jonsson.2007.V177.p597} was reviewed  by~\citet{FroeseFischer.2016.V49.p182004}. This method is also described in our recent papers~\citep{Wang.2018.V235.p27,Song.2020.V247.p70}. For this reason, only  computational procedures are described below.

In our MCDHF calculations, the multireference (MR) sets for even and odd parities include

even configurations:
$3p^4 3d$, $3s 3p^2 3d^2$, $3s 3p^3 4p$,  $3s 3p^4$, $3s^2 3d^3$, $3s^2 3p^2 3d$, $3s^2 3p^2 4d$, $3s^2 3p^2 4s$;

odd  configurations:
$3p^3 3d^2$, $3p^5$, $3s 3p 3d^3$,  $3s 3p^3 3d$, $3s 3p^3 4s$, $3s^2 3p 3d^2$, $3s^2 3p^2 4f$, $3s^2 3p^2 4p$, $3s^2 3p^3$.

By allowing single and double  substitutions from the $n=3,4$ electrons of the MR sets to orbitals with  $n\leq8, l\leq6$, and allowing single excitations of the $n=2$ electrons to orbitals with $n\leq6, l\leq4$,  configuration state function (CSF) expansions are generated. 
No substitutions are allowed from the $1s$ shell, which is defined as an inactive closed core.  
For both energy separations and  transition probabilities, the neglected electron correlations from $n=1,2$ are comparatively unimportant~\citep{Wang.2018.V235.p27,Song.2020.V247.p70}.
 
In the following RCI calculation, the transverse electron-photon interaction in the low-frequency limit and the  leading quantum electrodynamic (QED) effects (vacuum polarization and self-energy) corrections are included. 
In the final  CSF expansions for the different $J$ symmetries,  the number of CSFs  is, respectively, about 5.9 millions for even parity and 8.1 millions for odd parity. 

By using the $jj$-$LSJ$ transformation approach\\~\citep{Gaigalas.2017.V5.p6, Gaigalas.2004.V157.p239}, the  $jj$-coupled CSFs are transformed into $LSJ$-coupled CSFs, from which the $LSJ$ labels used by experimentalists are obtained.
\subsection{MBPT}\label{Sec:MBPT}
The MBPT method~\citep{Lindgren.1974.V7.p2441} is implemented in the FAC code by~\citet{Gu.2008.V86.p675,Gu.2007.V169.p154}. This method was used in our recent papers
	%~\citep{Wang.2014.V215.p26,Wang.2015.V218.p16,Wang.2016.V223.p3,Wang.2016.V226.p14}, 
	~\citep{Wang.2014.V215.p26,Wang.2015.V218.p16,Wang.2016.V223.p3,Wang.2016.V226.p14,Wang.2017.V229.p37,Wang.2018.V239.p30} to provide high accuracy atomic data for L- and M- shell ions.
In the  MBPT method, the Hilbert space of the full Hamiltonian is divided into two parts, i.e., a model space $M$ and an orthogonal space $N$.  We included all the CSFs of the MR set of the above MCDHF calculation in the $M$ space. All the possible CSFs generated by permitting single  and double  substitutions from the electrons of the MR sets are included in the $N$ space.  With the maximum $l$ value of 20, for single and double  excitations the maximum $n$ values are, respectively, 125 and 65.  The configuration interaction effects in the model space $M$ are considered in the self-consistent field calculations non-perturbatively.  The effects involving both spaces $M$ and $N$ are considered by using the second-order perturbation method. 

\section{Results and Discussions}~\label{results}
\subsection{Excitation energies}
In Table~\ref{table1}, excitation energies and lifetimes for the lowest 150 levels of the $3s^2 3p^3$, $3s 3p^4$, $3s^2 3p^2 3d$, $3s 3p^3 3d$, $3p^5$, and $3s^2 3p 3d^2$ configurations in \mbox{Ge XVIII} from our MCDHF calculations are provided. Excitation energies from the present MBPT calculations, as well as the energy differences between the MBPT and MCDHF results, are also included in this table.

In Table~\ref{table3}, the present MCDHF and MBPT excitation energies are compared with experimental values compiled in the NIST Atomic Spectra Database (ASD)~\citep{Kramida.2018.V.p}.
Calculated values obtained by~\citet{Hu.2020.V137.p1141} using the MCDHF method, and~\citet{Vilkas.2004.V37.p4763} using the
multireference M{\o}ller-Plesset (MR-MP) perturbation theory are also included in the table. 
As shown in Table~\ref{table3}, the previous calculations~\cite{Hu.2020.V137.p1141,Vilkas.2004.V37.p4763} were focused on the 41 low-lying states of the $3s^2 3p^3$, $3s 3p^4$, and $3s^2 3p^2 3d$  configurations. The average difference with the standard deviation, derived based on formulas (3-4) in~\cite{Wang.2017.V229.p37}, between computed excitation energies and the NIST experimental values are $229 \pm 446$ cm$^{-1}$ for our MCDHF calculations, $-610 \pm 660$ cm$^{-1}$ for our MBPT calculations,  $106 \pm 494$ cm$^{-1}$ for ~\citet{Vilkas.2004.V37.p4763}, and $1359 \pm 675$ cm$^{-1}$ for ~\citet{Hu.2020.V137.p1141},  respectively. Comparing with the previous MCDHF calculated values reported by~\citet{Hu.2020.V137.p1141}, there is generally a better agreement between our MCDHF values and the  NIST experimental values due to the larger extent of  electron correlation effects included in the present MCDHF calculations.
For all 41 states of the $3s^2 3p^3$, $3s 3p^4$, and $3s^2 3p^2 3d$  configurations, the average difference from the present MCDHF  excitation energies is $-130 \pm 412$ cm$^{-1}$ for MR-MP, and $-930 \pm 285$ cm$^{-1}$ for MBPT.

Looking at higher-lying levels (above the level with the key 41), good agreement between our two data sets (MCDHF and MBPT) is obtained. The energy difference in cm$^{-1}$ between our two calculations for each level is shown in Table~\ref{table1}. The average difference with standard deviation is found to be $-1127 \pm 446$ cm$^{-1}$, corresponding to the average relative difference  of $-0.09~\% \pm 0.03~\%$. The present MCDHF and MBPT results support each other, and certify the reliability of our calculations.

\subsection{Wavelengths, transition rates, and lifetimes}\label{sec_tr}
Wavelengths $\lambda$ and radiative transition data, including transition
rates $A$, weighted oscillator strengths $gf$,  line strength $S$, and branching fractions  (${\rm BF}_{ji} = A_{ji}/ \sum \limits_{k=1}^{j-1} A_{jk}$)  for E1, M1, electric quadrupole (E2), and magnetic quadrupole (M2) transitions among all the lowest 150 levels from the present MCDHF calculations  are reported in Table~ \ref{table2}. 
E1 and E2 radiative transition data are given in both length ($l$) and velocity ($v$) forms. Only transitions with BF  $\geq 10^{-5}$ are presented in this table.

Using the uncertainty estimation approach~\citep{Kramida.2013.V63.p313,Kramida.2014.V212.p11}, for E1 and E2 transitions we provide  the estimated uncertainties of line strengths $S$ adopting the NIST ASD~\citep{Kramida.2018.V.p} terminology (AA $\leq$ 1~\%, A$^{+}$ $\leq$ 2~\%, A $\leq$ 3~\%, B$^{+}$ $\leq$ 7~\%, B $\leq$ 10~\%, C$^{+}$  $\leq$ 18~\%,  C $\leq$ 25~\%,  D$^{+}$ $\leq$ 40~\%, D $\leq$ 50~\%, and E $>$ 50~\% ) in the last column of Table~ \ref{table2}.   
The difference $\delta S$ between line strengths $S_l$ and $S_v$ (in length and velocity forms respectively) is defined as $\delta S$ = $\left|S_{v}  - S_{l} \right|$/$\min (S_{v}$,~$S_{l})$. The averaged uncertainties $\delta S_{av}$ for line strengths $S$  for E1 transitions in various ranges of $S$ are assessed to be 0.2~\% for $S \geq 10^{0}$; 0.5~\% for $10^{0} > S \geq 10^{-1}$; 1.1~\% for $10^{-1} > S \geq 10^{-2}$; 2.4~\% for $10^{-2} > S \geq 10^{-3}$; 4.7~\% for $10^{-3} > S \geq 10^{-4}$; 8~\% for $10^{-4} > S \geq 10^{-5}$, and 35~\% for $10^{-5} > S \geq 10^{-6}$. Then, the larger of  $\delta S_{av}$ and $\delta S_{ji}$ is accepted as the uncertainty of each particular line strength. 
In Table~ \ref{table2}, about 8~\% of E1 $S$ values  have uncertainties of  $\leq$ 1~\% (AA), 27~\% have uncertainties of  $\leq$ 2~\% (A+), 
25~\% have uncertainties of  $\leq$ 3~\% (A), 
24~\% have uncertainties of  $\leq$ 7~\% (B+), 
11~\% have uncertainties of  $\leq$ 10~\% (B), 
4~\% have uncertainties of  $\leq$ 18~\% (C+), 
0.6~\% have uncertainties of  $\leq$ 25~\% (C), 
and 1~\% have uncertainties of  $\leq$ 40~\% (D+), 
while only 0.2~\% have uncertainties of  $>$ 40~\% (D and E). 
In the spirit of the uncertainty estimation approach~\citep{Kramida.2013.V63.p313,Kramida.2014.V212.p11}, the estimated uncertainties of line strengths $S$ for E2  transitions are also estimated and  listed in Table~\ref{table2}.

Among different previous calculations~\cite{Hu.2020.V137.p1141,Vilkas.2004.V37.p4763,Charro.2000.V33.p1753,Fritzsche.1998.V68.p149,Huang.1984.V30.p313,Fischer.1986.V19.p137}, radiative rates for the M1 transitions within the lowest 5 levels of the $3s^2 3p^3$ configuration were provided by~\citet{Huang.1984.V30.p313} using the MCDHF method, and ~\citet{Fischer.1986.V19.p137} using  the multi-configuration Hartree-Fock (MCHF) method. In Table~\ref{tableM1}, our MCDHF values are compared with previous results~\cite{Fischer.1986.V19.p137,Huang.1984.V30.p313}. Comparing with the previous MCHF radiative rates, the MCDHF results reported by~\citet{Huang.1984.V30.p313} show a better agreement with our MCDHF values. The difference from the present MCDHF values for those M1 transitions is a few percent for the previous MCDHF results obtained by~\citet{Huang.1984.V30.p313}.

Our MCDHF radiative lifetimes $\tau_{\rm MCDHF}^l$ (in s) in the length  form and $\tau_{\rm MCDHF}^v$ (in s)  in the velocity form for the lowest 150 levels of the $3s^2 3p^3$, $3s 3p^4$, $3s^2 3p^2 3d$, $3s 3p^3 3d$, $3p^5$, and $3s^2 3p 3d^2$ configurations in \mbox{Ge XVIII}, which are calculated by considering all possible E1, E2, M1, and M2 transitions, are provided in Table~\ref{table1}. Our MCDHF radiative lifetimes $\tau_{\rm MCDHF}^l$ and  $\tau_{\rm MCDHF}^v$ show good agreement. The average deviation between $\tau_{\rm MCDHF}^l$ and  $\tau_{\rm MCDHF}^v$ for all 150 levels is 1~\%.

\section{Conclusions}~\label{conclusions}
Using the MCDHF method combined with the RCI approach, including the transverse electron-photon interaction in the low-frequency limit and the  leading QED   corrections,  calculations have been performed  for the lowest 150 levels of the $3s^2 3p^3$, $3s 3p^4$, $3s^2 3p^2 3d$, $3s 3p^3 3d$, $3p^5$, and $3s^2 3p 3d^2$ configurations in \mbox{Ge XVIII}.
Excitation energies, radiative transition data, and  lifetimes  are reported. 

The accuracy of energy levels from the MCDHF calculations is estimated by comparing the MCDHF results with available experimental data.  The difference from the NIST experimental values  is $-229 \pm 446$ cm$^{-1}$ for the MCDHF excitation energies of the lowest 41 levels. 
To assess the accuracy of the MCDHF excitation energies involving higher-lying levels, we  also perform  calculations using the MBPT method, and provide excitation energies. The average absolute difference of the present MBPT and MCDHF energy values is $-1127 \pm 446$ cm$^{-1}$ for high-lying levels (above the level with the key 41), corresponding to the average relative difference  of $0.09~\%\pm0.03~\%$, where the standard deviations are indicated after the values. 
The uncertainty of the  line strength  is assessed  for each transition, and is available in Table~\ref{table2}. 
The resulting accurate and consistent MCDHF data set will be useful for  line identification and modeling purposes, which can also be considered as a benchmark for other calculations.

\section{Acknowledgments}
We acknowledge the support from the National Key Research and Development Program of China under Grant No.~2017YFA0403200,  the Science Challenge Project of
China Academy of Engineering Physics (CAEP) under
Grant No.~TZ2016005, the National Natural Science Foundation of China (Grant No.~11703004, No.~12074081, and No.~U2031135), the Natural Science Foundation of Hebei Province, China (A2019201300), and the Natural Science Foundation of Educational Department of Hebei Province, China (BJ2018058). 
Kai Wang expresses his gratitude to the support from the visiting researcher program of  Fudan University.

\onecolumn

%\section*{References}
\bibliographystyle{model}
\bibliography{ref}
%\bibliography{../../../../../article/ref}

%\include{table3.tex}
\onecolumn

\onecolumn

\clearpage

\section*{Tables}

\linespread{1}
\tiny
\setlength{\tabcolsep}{5pt}
\begin{longtable}{clrrrrrrl}
	\caption{\label{table1}Excitation energies (in cm$^{-1}$) and radiative lifetimes (in s) 
		for the lowest 150 levels of the $3s^2 3p^3$, $3s 3p^4$, $3s^2 3p^2 3d$, $3s 3p^3 3d$, $3p^5$, and $3s^2 3p 3d^2$ configurations in \mbox{Ge XVIII}.
		$E_{\rm MCDHF}$ - the MCDHF excitation energies;  $E_{\rm MBPT}$ - the MBPT excitation energies; $\Delta E$ - the differences  (in cm$^{-1}$)   between the  MCDHF  and MBPT excitation energies;
		$\tau_{\rm MCDHF}^l$ - the MCDHF lifetimes in length form;   $\tau_{\rm MCDHF}^v$ - the MCDHF lifetimes  in velocity form;  $LS$-composition -  the $LS$ eigenvector compositions from the MCDHF calculations. }\\
	\toprule
	Key & Level& $J\pi$ &$E_{\rm MCDHF}$ &$E_{\rm MBPT}$ &$\Delta E$& $\tau_{\rm MCDHF}^l$& $\tau_{\rm MCDHF}^v$& $LS$-composition\\
	\midrule
	\endfirsthead
	\caption{(\emph{continued})} \\
	\toprule
	Key & Level& $J\pi$ &$E_{\rm MCDHF}$ &$E_{\rm MBPT}$ &$\Delta E$& $\tau_{\rm MCDHF}^l$& $\tau_{\rm MCDHF}^v$& $LS$-composition\\
		\midrule
	\endhead
	\endfoot
	\bottomrule
	\endlastfoot
1   & $3s^{2}\,3p^{3}(^{4}_{3}S)~^{4}S_{3/2}^{\circ}$                                            &  3/2-& 0       &0       &      &          &           & 0.83 + 0.11~$3s^{2}\,3p^{3}(^{2}_{1}P)~^{2}P^{\circ}$ + 0.03~$3s^{2}\,3p^{3}(^{2}_{3}D)~^{2}D^{\circ}$                                                                                                   \\ 
2   & $3s^{2}\,3p^{3}(^{2}_{3}D)~^{2}D_{3/2}^{\circ}$                                            &  3/2-& 56391   &56045   & -346 & 7.459E-04&  7.459E-04& 0.72 + 0.15~$3s^{2}\,3p^{3}(^{2}_{1}P)~^{2}P^{\circ}$ + 0.09~$3s^{2}\,3p^{3}(^{4}_{3}S)~^{4}S^{\circ}$                                                                                                   \\ 
3   & $3s^{2}\,3p^{3}(^{2}_{3}D)~^{2}D_{5/2}^{\circ}$                                            &  5/2-& 73792   &73462   & -330 & 5.676E-03&  5.677E-03& 0.97                                                                                                                                                                                                     \\ 
4   & $3s^{2}\,3p^{3}(^{2}_{1}P)~^{2}P_{1/2}^{\circ}$                                            &  1/2-& 112610  &112242  & -368 & 2.884E-04&  2.884E-04& 0.96                                                                                                                                                                                                     \\ 
5   & $3s^{2}\,3p^{3}(^{2}_{1}P)~^{2}P_{3/2}^{\circ}$                                            &  3/2-& 142533  &142198  & -335 & 1.191E-04&  1.191E-04& 0.70 + 0.22~$3s^{2}\,3p^{3}(^{2}_{3}D)~^{2}D^{\circ}$ + 0.05~$3s^{2}\,3p^{3}(^{4}_{3}S)~^{4}S^{\circ}$                                                                                                   \\
\end{longtable}
\clearpage

\newpage
\linespread{1}
\scriptsize
\setlength{\tabcolsep}{5pt}
\begin{longtable}{llccccccccc}
	\caption{\label{table3}Energy levels (in cm$^{-1}$) relative to the ground state for the lowest 41 states of the $3s^2 3p^3$, $3s 3p^4$, and $3s^2 3p^2 3d$  configurations. NIST -- experimental values from the NIST ASD~\citep{Kramida.2018.V.p}; MCDHF --  present MCDF results; MBPT -- present MBPT results; Vilkas -- theoretical values from \citet{Vilkas.2004.V37.p4763}. Hu -- theoretical values from \citet{Hu.2020.V137.p1141};   $\Delta E$ -- differences  (in cm$^{-1}$)  between calculated energies from different calculations and the NIST values.}\\
	\midrule
	Key & State & NIST & \multicolumn{2}{c}{MCDHF} &  \multicolumn{2}{c}{MBPT}  & \multicolumn{2}{c}{Vilkas}& \multicolumn{2}{c}{Hu}\\
\cmidrule(r){4-5}\cmidrule(r){6-7}\cmidrule(r){8-9}\cmidrule(r){10-11}
&       &   $E$  &  $E$                       &  $\Delta E$           & $E$  &  $\Delta E$ & $E$ & $\Delta E$& $E$ & $\Delta E$ \\
	\midrule
	\endfirsthead
	\caption{(\emph{continued})} \\
	\toprule
	Key & State & NIST & \multicolumn{2}{c}{MCDHF} &  \multicolumn{2}{c}{MBPT}  & \multicolumn{2}{c}{Vilkas}& \multicolumn{2}{c}{Hu}\\
	\cmidrule(r){4-5}\cmidrule(r){6-7}\cmidrule(r){8-9}\cmidrule(r){10-11}
	&       &   $E$  &  $E$                       &  $\Delta E$           & $E$  &  $\Delta E$ & $E$ & $\Delta E$& $E$ & $\Delta E$ \\
	\midrule
	\endhead
	\midrule
	\endfoot
	\bottomrule
	\endlastfoot
1   & $3s^{2}\,3p^{3}(^{4}_{3}S)~^{4}S_{3/2}^{\circ}$                                            &0       &0      &     &0       &      &0      &      &0       &    \\ 
2   & $3s^{2}\,3p^{3}(^{2}_{3}D)~^{2}D_{3/2}^{\circ}$                                            &56117   &56391  & 274 &56045   &-72   &56172  &55    &57077   &960 \\ 
3   & $3s^{2}\,3p^{3}(^{2}_{3}D)~^{2}D_{5/2}^{\circ}$                                            &73646   &73792  & 146 &73462   &-184  &73601  &-45   &74606   &960 \\ 
4   & $3s^{2}\,3p^{3}(^{2}_{1}P)~^{2}P_{1/2}^{\circ}$                                            &112334  &112610 & 276 &112242  &-92   &112351 &17    &113963  &1629\\ 
5   & $3s^{2}\,3p^{3}(^{2}_{1}P)~^{2}P_{3/2}^{\circ}$                                            &142132  &142533 & 401 &142198  &66    &142344 &212   &143685  &1553\\ 
6   & $3s~^{2}S\,3p^{4}(^{3}_{2}P)~^{4}P_{5/2}$                                                  &        &407960 &     &407218  &      &407303 &      &407243  &    \\ 
7   & $3s~^{2}S\,3p^{4}(^{3}_{2}P)~^{4}P_{3/2}$                                                  &        &437196 &     &436516  &      &436574 &      &436471  &    \\ 
8   & $3s~^{2}S\,3p^{4}(^{3}_{2}P)~^{4}P_{1/2}$                                                  &        &446553 &     &445870  &      &445911 &      &445910  &    \\
\end{longtable}
\clearpage

\clearpage
\linespread{1}
\tiny
\setlength{\tabcolsep}{5pt}
\begin{longtable}{llcccccccccl}
	\caption{\label{table2}Transition wavelengths $\lambda$ (in {\AA}), transition rates $A$ (in s$^{-1}$), weighted oscillator strengths $gf$, and line strengths $S$ (in a.u.) 
		between the lowest 150 states  listed in Table~\ref{table1}. E1 and E2 transition data in both the length ($l$) and velocity ($v$) forms are provided. Type is the type of the multipole, and BF is the branching fraction from the upper level. The last column (Acc.) represents the estimated accuracy of the $S$ values using the terminology of the NIST ASD. Only transitions with BF  $\geq 10^{-5}$ are presented.}\\
	\toprule
	$i$ & $j$& 		$\lambda$ & Type  &BF & $A^l$ & $gf^l$ & $S^l$ & $A^v$ & $gf^v$ & $S^v$&  Acc.     \\	\midrule
	\endfirsthead
	\caption{(\emph{continued})} \\
	\toprule
	$i$ & $j$& 		$\lambda$ & Type  &BF & $A^l$ & $gf^l$ & $S^l$ & $A^v$ & $gf^v$ & $S^v$&  Acc.     \\	
\midrule
	\endhead
	\midrule
	\endfoot
	\bottomrule
	\endlastfoot
    1&    2&  1.77332E+03&  M1&  9.998E-01&  1.340E+03&  2.528E-06&  1.108E+00&           &           &           &       \\
1&    2&  1.77332E+03&  E2&  2.142E-04&  2.871E-01&  5.415E-10&  1.798E-02&  2.819E-01&  5.317E-10&  1.766E-02&  B+   \\
1&    3&  1.35516E+03&  M1&  7.513E-01&  1.324E+02&  2.187E-07&  7.328E-02&           &           &           &       \\
1&    3&  1.35516E+03&  E2&  8.368E-03&  1.474E+00&  2.436E-09&  3.610E-02&  1.448E+00&  2.392E-09&  3.546E-02&  B+   \\
2&    3&  5.74697E+03&  M1&  2.404E-01&  4.235E+01&  1.258E-06&  1.788E+00&           &           &           &       \\
1&    4&  8.88019E+02&  M1&  8.519E-01&  2.954E+03&  6.984E-07&  1.534E-01&           &           &           &       \\
1&    4&  8.88019E+02&  E2&  1.162E-03&  4.030E+00&  9.530E-10&  3.975E-03&  3.996E+00&  9.448E-10&  3.940E-03&  D    \\
2&    4&  1.77876E+03&  M1&  1.459E-01&  5.058E+02&  4.798E-07&  2.111E-01&           &           &           &       \\
\end{longtable}
%\begin{enumerate}[]
%	\item Only transitions among the 41 low-lying  states of the $3s^2 3p^3$, $3s 3p^4$, and $3s^2 3p^2 3d$  configurations are shown here. Table~\ref{table2} is available online in its entirety at the \emph{JQSRT} website.
%\end{enumerate}
\clearpage

\newpage
\linespread{1}
\tiny
\setlength{\tabcolsep}{5pt}
\begin{longtable}{lllllccccccccc}
	\caption{\label{tableM1} Comparison of radiative rates for the M1 transitions within the lowest 5 levels of the $3s^2 3p^3$ configuration. MCDHF -- present MCDHF results; Huang -- theoretical values from \citet{Huang.1984.V30.p313}. Froese Fischer -- theoretical values from \citet{Fischer.1986.V19.p137}.}\\
	\toprule
	$i$ & $j$  & Lower level & Upper level & Type & \multicolumn{3}{c}{$A$}  & \multicolumn{2}{c}{ Percentage difference for  $A$ }& \\
	\cmidrule(r){6-8}\cmidrule(r){9-11}
	& & & &  & MCDHF & Huang &   Froese Fischer & Huang \& MCDHF & Froese Fischer \& MCDHF \\
	\midrule
	\endfirsthead
	\caption{(\emph{continued})} \\
	\toprule
	
	\midrule
	\endhead
	\midrule
	\endfoot
	\bottomrule
	\endlastfoot
	1	 &  2&	$3s^{2}\,3p^{3}(^{4}_{3}S)~^{4}S_{3/2}^{\circ}$	&  $3s^{2}\,3p^{3}(^{2}_{3}D)~^{2}D_{3/2}^{\circ}$&  M1  & 1.340E+03 &   1.440E+03 & 1.50E+03 & 7  &  12   \\
	1	 &  3&	$3s^{2}\,3p^{3}(^{4}_{3}S)~^{4}S_{3/2}^{\circ}$	&  $3s^{2}\,3p^{3}(^{2}_{3}D)~^{2}D_{5/2}^{\circ}$&  M1  & 1.324E+02 &   1.354E+02 & 1.37E+02 & 2  &  3    \\
	2	 &  3&	$3s^{2}\,3p^{3}(^{2}_{3}D)~^{2}D_{3/2}^{\circ}$	&  $3s^{2}\,3p^{3}(^{2}_{3}D)~^{2}D_{5/2}^{\circ}$&  M1  & 4.235E+01 &   4.585E+01 & 5.80E+01 & 8  &  37   \\
	1	 &  4&	$3s^{2}\,3p^{3}(^{4}_{3}S)~^{4}S_{3/2}^{\circ}$	&  $3s^{2}\,3p^{3}(^{2}_{1}P)~^{2}P_{1/2}^{\circ}$&  M1  & 2.954E+03 &   3.084E+03 & 2.81E+03 & 4  &  -5   \\
	2	 &  4&	$3s^{2}\,3p^{3}(^{2}_{3}D)~^{2}D_{3/2}^{\circ}$	&  $3s^{2}\,3p^{3}(^{2}_{1}P)~^{2}P_{1/2}^{\circ}$&  M1  & 5.058E+02 &   5.374E+02 & 4.69E+02 & 6  &  -7   \\
	1	 &  5&	$3s^{2}\,3p^{3}(^{4}_{3}S)~^{4}S_{3/2}^{\circ}$	&  $3s^{2}\,3p^{3}(^{2}_{1}P)~^{2}P_{3/2}^{\circ}$&  M1  & 2.532E+03 &   2.656E+03 & 2.50E+03 & 5  &  -1   \\
	2	 &  5&	$3s^{2}\,3p^{3}(^{2}_{3}D)~^{2}D_{3/2}^{\circ}$	&  $3s^{2}\,3p^{3}(^{2}_{1}P)~^{2}P_{3/2}^{\circ}$&  M1  & 4.499E+03 &   4.468E+03 & 4.04E+03 & -1 &  -10  \\
	3	 &  5&	$3s^{2}\,3p^{3}(^{2}_{3}D)~^{2}D_{5/2}^{\circ}$	&  $3s^{2}\,3p^{3}(^{2}_{1}P)~^{2}P_{3/2}^{\circ}$&  M1  & 1.176E+03 &   1.165E+03 & 9.92E+02 & -1 &  -16  \\
\end{longtable}

\end{document}